# Driven Morse Oscillator: Model for Multi-photon Dissociation of Nitrogen Oxide

By Julian Juhi-Lian Ting

Institute of Physics, Tsing-Hua University, Hsin-Chu, Taiwan 30043, Republic of China †

Within a one-dimensional semi-classical model with a Morse potential the possibility of microwave multi-photon dissociation of vibrationally excited nitrogen oxide was studied. The dissociation thresholds of typical driving forces and couplings were found to be similar, which indicates that the results were robust to variations of the potential and the definition of dissociation rate.

† E-mail address: jlting@phys.nthu.edu.tw

## 1. Introduction

Experimental microwave multi-photon dissociation of nitrogen oxide (NO) is not yet possible, while the corresponding optical photodissociation is easily observed via electronic transitions. However, it is important to know the possibility and how one can make it, firstly because of the abundance of NO in nature and living bodies, and secondly because the intensity of lasers has reached the order of magnitude needed to do the experiment. Although hydrogen fluoride (HF) has a smaller dissociation energy, from an experimental view, because of the stability, simplicity and well documented properties of NO, easier preparation of intense molecular beams of NO than for HF, easier efficient probing, NO has been chosen for investigation instead of HF. Years ago experiments were conducted mostly at the ground state of NO prepared by supersonic expansion, recently it has become possible to transfer population efficiently to other higher states: Yang *et al* (1990) report experiments have been performed to populate NO molecules into an initial vibrational state as great as 25 by stimulated emission pumping whereas Schiemann *et al* (1993) showed recently populate to a vibrational state 6 by a stimulated Raman process involving adiabatic passage (STIRAP) method using pulse laser. In our study we choose parameters for NO whenever molecular properties are needed for our calculations, so that future experiments can easily compare with the results presented here. The behaviour of NO is expected to be typical of that of most other diatomic molecules.

Theoretically much previous work concentrated on the corresponding atomic systems. In particular, Blümel and Smilansky (1987) did a calculation similar to the present work for the hydrogen atom and Broeckhove (1992) did a calculation for HF. But no previous calculations for the dissociation thresholds for every initial eigenstates has been performed. In this paper the Morse potential-energy function was used for the inter-atomic interaction, however only the vibrational levels are considered because Morse (1929), Pekeris (1946) have pointed out that the rotational influence on vibration can be taken into account by suitably adjusting the Morse parameters.

The Morse potential has been used by Lènnard-Jones and Strachan (1935), Strachan (1935) to study the interaction of atoms and molecules with solid surfaces in the early days. Isnor and Richardson (1971), Walker and Preston (1977) have been investigated the driven Morse oscillator as a model for infrared multi-photon excitation and dissociation of molecules, stimulated by the possibility of laser isotope separation and bond-selective chemistry. The case for the Morse potential here lacks the complication of infinitely bound eigenstate and the singularity of the Coulomb potential at the origin for the hydrogen atom.

The methods used for calculating multi-photon dissociation generally divided into non-linear classical mechanics methods like Goggin and Milonni (1988) and semi-quantum methods like Heather and Metiu (1987), Heather and Metiu (1988), Tanner and Maricq (1988), Tanner and Maricq (1989). The later approach is used in this study.

## 2. Model

We consider an isolated non-rotating NO molecule interacting with a plane-polarised harmonic laser field. The dimensionless Hamiltonian of a free Morse oscillator for a diatomic molecule is

$$H_0 = \frac{p^2}{2} + \frac{(1-e^{-z})^2}{2}, \tag{1}$$

| parameter | symbol | value |
|---|---|---|
| range parameter of the potential | $\alpha$ | $2.7675 \times 10^8 cm^{-1}$ |
| reduced mass | $m$ | $7.4643u$ |
| dissociation energy | $D$ | $6.4968 eV$ |
| Morse frequency | $\omega_0 = \sqrt{2D\alpha^2/m}$ | $1904.2 cm^{-1}$ |

TABLE 1. Typical parameters for NO.

in which $z = \alpha(r - r_e)$ denotes displacement of inter-atomic distance from equilibrium, $p = p_z/\sqrt{2mD}$, and $p_z$ is the momentum conjugate to $z$. The parameters for NO are given by Huber and Herzberg (1979) and summarised in the TABLE. The Hamiltonian for the Morse oscillator in the presence of a typical harmonic laser field reads

$$H = H_0 - \frac{A\Omega}{2}\mu(z)\Phi(\Omega t), \qquad (2)$$

in which $\mu$ is the dipole moment operator. In the formula above two dimensionless variables are introduced, namely $\Omega = \omega_L/\omega_0$ with $\omega_L$ the laser frequency, and $A = qE_L/\alpha\Omega D$ is the dimensionless field strength with $E_L$ the field strength of the laser and $q$ is the effective dipole charge of the molecule. In the numerical calculations below various types of couplings are considered.

The discrete eigenfunctions for the free Morse oscillator were given by Morse (1929). We denote the eigenfunctions of $H_0$ by $|n>$ with the corresponding dimentionless eigenvalues equal to

$$E_n = \frac{(n + \frac{1}{2})}{\lambda} - \frac{(n + \frac{1}{2})^2}{2\lambda^2}, \qquad (3)$$

in which $\lambda = \sqrt{2mD}/\alpha\hbar = 55.04$. The term multi-photon is generally applied to this system, as each photon has energy about $0.24 eV$ near the Morse frequency ($\omega_0$), whereas the energy difference between the ground state and the first excited state of the free Morse oscillator is about $0.23 eV$; about 28 photons must be absorbed for the transition from the ground state to the continuum.

Our main concern is the transition between each excited state of the free Morse oscillator and dissociation. It is possible that with various initial states $|n>$, driving frequencies and driving amplitudes the dissociation period required might change dramatically. The problem of this calculation is whether our model can really represent the molecular system for which we intend it for and what is our definition of dissociation. Does it represent the real situation and how robust are the results? To address these questions we present two typical definitions of dissociation and driving forces.

## 3. Numerical results

The following numerical results were obtained using methods described and verified previously by Ting *et al* (1992), which is a fast Fourier-transformed grid method also considered by Feit *et al* (1982), Leforestier *et al* (1991).

### 3.1. *Dipole coupling*

Following Walker and Preston (1977), Goggin and Milonni (1988) on a driven Morse oscillator, as a consistency check in the first instance, a cosinusoidal wave form was used.

4Julian Juhi-Lian Ting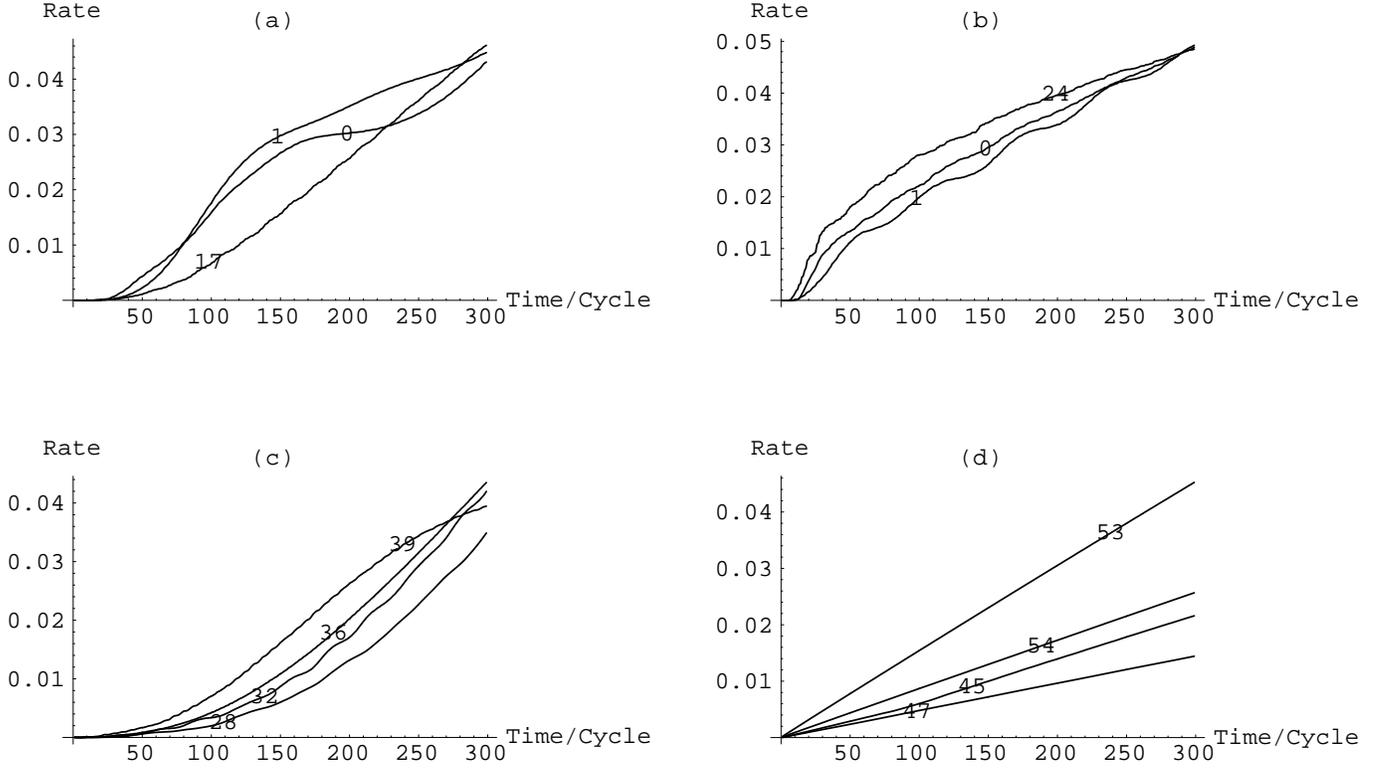

FIGURE 1. Dissociation histories of four types observed at critical amplitudes $A = A_c$ and laser frequency $\Omega = 0.9$; (a) the curves rises smoothly; (b) the curves rises sharply; (c) curves which are distinct from those of neighbouring states'; (d) straight lines.

Therefore the Hamiltonian in Eq. (2) reads

$$H = H_0 - \frac{A\Omega}{2} z \cos(\Omega t). \qquad (4)$$

According to Goggin and Milonni (1988), Chelkowski and Bandrauk (1990) in the study of quantum chaos the definition of dissociation rate is

$$P_{diss}(t) = 1 - \sum_n |<n|\psi(t)>|^2. \qquad (5)$$

We *define* $P_{diss}(t) = 0.05$ to be dissociated. If the molecule cannot reach such a dissociation rate within 300 cycles we consider it to be not dissociable under such conditions of field intensity and frequency. This definition is almost equivalent to comparing the dissociation rate after 300 cycles. According to Leforestier *et al* (1991) the latter should be easier to calculate. However, they are distinct. Furthermore, the definition of $P_{diss}$ is only approximate because, when the field is applied, the projection onto the bound states does not represent the true bound-state population unless one assumes that the electro-magnetic field can be terminated instantaneously. The number 300 and 0.05 are set according to previous experience although there are some guiding principles: 300 cycles corresponds to a pulse duration about 5 picosecond, which is the time scale for bound dissociation to occur and these choices are analogous to the atomic ionization calculation devised by Blümel and Smilansky (1987).

The dissociation histories from Eq. (5) were calculated at $\Omega = 0.9$. Fig.1 shows disso-ciation histories of four types at critical amplitudes, i.e., the largest amplitude at which the molecule just fails to be dissociated within 300 vibrations. There are several such critical amplitudes for some low lying initial states ( *vide infra* ). This classification is not rigourous, but it represents at least some typical dissociation histories. It is a result of investigating about 500 curves obtained for the calculation for Fig.3. Ting *et al* (1992) showed some other curves. A brief description for each type follows: For the first type the curve rises smoothly,while for the second type the curve rises sharply. We noticed for those initial states with two transitions the larger value of $A_c$ corresponds to the second type whereas the smaller value corresponds to the first type. We suspect that all those states belonging to the second type should have a smaller value of $A_c$, although only part of them were found. The third type is essentially the same as the first type but their neighbours belong to the second type. The fourth type belongs to the initial states exceeding 44; the dissociation history is a straight line. Except for the latter type, for which the figure cannot display very clearly, the reason for a greater dissociation period is not only that the slope of the curve was changed, but rather that there were several flat regions in which there was very little molecular dissociation. As a further investigation for the third type, Fig.2 compares at the same amplitude with initial states near $n_0 = 28$, indicating that some initial states such as $n_0 = 28$ dissociate more readily than their neighbours. These states are the ionization windows proposed by Blümel and Smilansky (1987). There are strong indications that some states are easier to dissociate than others. The dissociation process might be taken as a two-step process; i.e., the population first transferred to those easy states then to other states. Generally the populations deployed themselves soon after the process began and stayed well localised, as Walker and Preston (1977) have observed. Shankar (1980), Pfeifer (1993) have discussed that an initial set-up time is required for any periodic system is a general property, because the system has to sense the periodic perturbation. Various quantities such as the inter-atomic distance for initial states and the average change of energy with time were also computed. A preliminary result about the inter-atomic distance shows that immediately after the field is switched on states with $n_0 > 25$ moves in different direction from $n_0 < 25$ away from the inter-atomic equilibrium position.

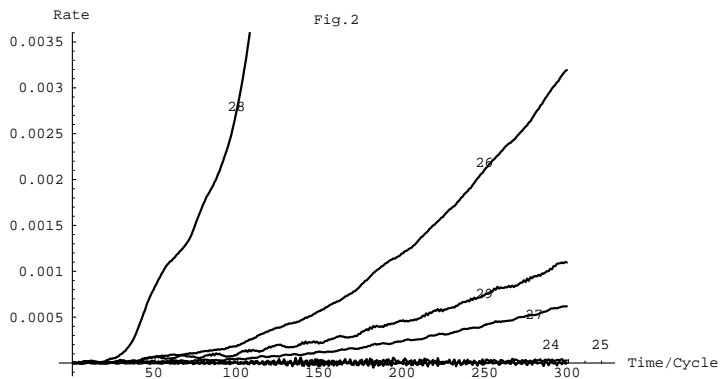

FIGURE 2. Dissociation histories for initial states from 24 to 29 at A=0.01 and $\Omega = 0.9$; $n_0 = 28$ dissociate more readily than its neighbouring states.

In Fig.3 the critical field required for the molecule to become dissociable is plotted. For multi-critical states the lowest ones are connected and the others marked. Therefore the plot is only a sufficient criterion for dissociation, not a necessary condition. The reason is that as we use numerical methods we can know only the points we computed. The amplitude resolution of our calculation is $\Delta A = 0.0005$. If there was a narrow transition less than 0.0005 one might not be able to find it. However if there exists such a point one might have difficulty to find experimentally because a laser of precise power would be implied. We further noted that in the present studies we find at most two critical amplitudes while in another study using two-color laser fields we find up to three ones. The multi-critical amplitude is also a result of our selection of critical dissociation periods as 300. However, it implies that decreased field strength does not necessarily increase the dissociation period. Lu *et al* (1991). noted that the finite size of $\hbar$ probably places a characteristic scale in parameter space below which the dissociation boundary ceases to show fractal structure. Another feature of the figure is that the largest value of $A_c$ is several orders of magnitude larger than the smallest one. A calculation shows for $n_0 \geq 6$ the photon energy becomes larger than the energy difference between adjacent states.

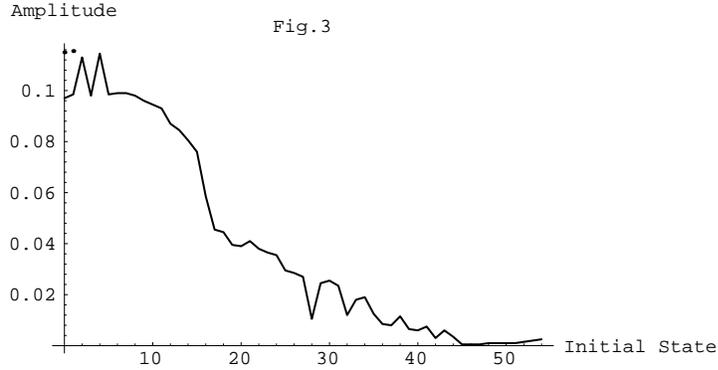

FIGURE 3. Critical amplitude ($A_c$) for various initial states at $\Omega = 0.9$; the Hamiltonian is Eq. (4) and the dissociation rate is Eq. (5); for low-lying states which have two $A_c$ only the lower one is connected.

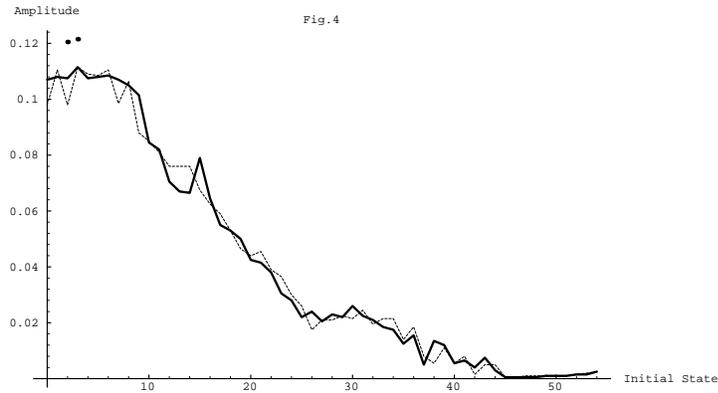

FIGURE 4. Critical amplitude ($A_c$) for various initial states at (a) $\Omega = 0.89$ (solid thick line) and (b) $\Omega = 0.91$ (thin dashed line); the Hamiltonian is Eq. (4) and the dissociation rate is Eq. (5).

To see whether $\Omega = 0.9$ is the most efficient frequency for all initial states we tested $\Omega = 0.89$ and $\Omega = 0.91$. The critical amplitudes are plotted in Fig.4. As we had insufficient computer time to run over all states and all frequencies, we selected $n_0 = 1, 17$ and 24 to see values of $A_c$ alter for frequencies from 0.8 to 1.1 at interval 0.01. The plots appear in Fig.5.

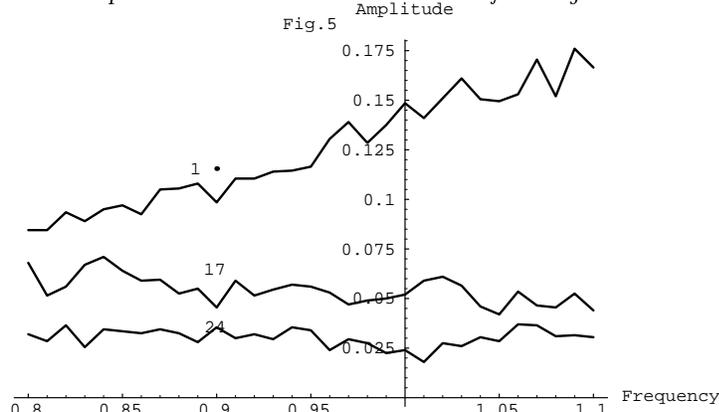

FIGURE 5. Critical amplitude ($A_c$) vs. frequency for initial states (a) $n_0 = 1$, (b) $n_0 = 17$ and (c) $n_0 = 24$.

### 3.2. *Exponential coupling*

The perturbation in Eq. (4), although simple and easily done by the Floquet method, has several problems in relation to reality. Firstly, the molecular dipole and laser coupling in equation Eq. (4) assumes a constant infinite range of coupling, which is satisfactory for an atomic system, but not for a real molecule. Secondly, the field should be zero at $t = 0$ and it should gradually increase from zero to a finite value; this is the so called adiabatic switching problem. Sakurai (1967) has noted part of the second point. However, it is rather mystical that cosinusoidal wave form passed down for generations. Therefore, another exponential form of coupling

$$H = H_0 - \frac{A\Omega}{2}(z+a)e^{-(z+a)/b}\sin(\Omega t)\sin^2\left(\frac{2\pi t}{T}\right), \tag{6}$$

was used to test whether there is any effect of the approximation. The problem of the definition of dissociation rate was pointed out above (below Eq.(5)); here we alter it to comply with Heather and Metiu (1987)

$$P_{diss}(t) = 1 - <\psi(t)|\psi(t)>. \tag{7}$$

This definition is appropriate if coherent excitation become important. The variables $a$ and $b$ were chosen to have the values 2 and 1 respectively in the calculation. Heather and Metiu (1987), Heather and Metiu (1988), Tanner and Maricq (1988), Tanner and Maricq (1989) have also used coupling functions of similar form. We take $T$, the pulse duration, to be 300 optical cycles. This choice has the advantage that, when $A = A_c$, $T_d$ is almost 300 cycles and the field strength is almost zero at that time. Hence there is little effect of the definition used for the dissociation rate. The resulting critical front is obtained in Fig.6 using a thick line. There is no essential distinction except a scaling constant that can be adjusted by choosing a and b. However, we noticed for smaller values of initial state there are more variations than for higher initial states. To answer whether $T$ is large enough a longer period $T = 600$ is tested. The result is plotted in Fig.6 with a thin dashed line, which shows very similar behaviour.

### 4. Summary

In conclusion, thresholds of vibrational dissociation of NO are presented in this paper. The main result of this paper is shown in Fig.6. Two perturbations and definitions of dissociation rate lead to similar curves with the transition from high $A_c$ to low $A_c$ shifted slightly to higher initial states. Several other kinds of laser fields and couplings were tested for which similar curves were obtained. Therefore we conclude that our results are robust to variation of the potential function as our main concern is the critical amplitude which is an indirect result. Variation of other types such as the choice of laser duration and definition of dissociation threshold are less significant. A rough estimation for the decrease of $A_c$ is given. Care must be given in comparing the results presented in this work with those from experiments. For instance, because of the finite cross section of the beam, not all molecules experience the same field strength.

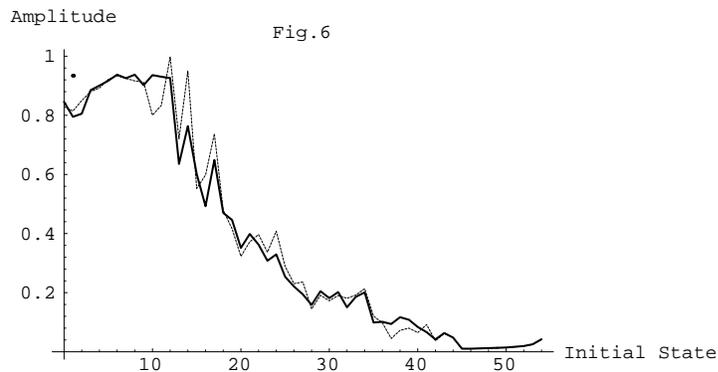

FIGURE 6. Critical amplitude ($A_c$) for various initial states and modified potential Eq. (6) at $\Omega = 0.9$. Thick line uses $T = 300$ whereas thin dashed line was obtained with $T = 600$.

## 5. Acknowledgement